\documentclass[aip,jap,amsfonts,amssymb,amsmath,groupedaddress,twocolumn]{revtex4}

\usepackage{graphicx}
\usepackage{amsmath}
\usepackage{amsfonts}
\usepackage{amssymb}
\usepackage{graphicx}
\usepackage{epstopdf}

\begin{document}

\title{Analysis of simple scattering models on the thermoelectric performance of analytical electron dispersions}

\author{Cameron Rudderham}
\affiliation{Department of Physics and Atmospheric Science, Dalhousie University, Halifax, Nova Scotia, Canada, B3H 4R2}
\author{Jesse Maassen}
\email{jmaassen@dal.ca}
\affiliation{Department of Physics and Atmospheric Science, Dalhousie University, Halifax, Nova Scotia, Canada, B3H 4R2}

\begin{abstract}
Recent first-principles electron-phonon scattering calculations of heavily-doped semiconductors suggest that a simple DOS scattering model, wherein the electronic scattering rates are assumed to be proportional to the density-of-states, better approximates the rigorous scattering characteristics compared to the commonly used constant relaxation-time and constant mean-free-path approximations. This work investigates how the thermoelectric properties predicted with the DOS model compare to the other two scattering models, using three analytical electron dispersions (parabolic band in 3D/2D/1D, Kane band in 3D/2D/1D, and ring-shaped quartic band in 2D). Our findings show that the scattering models can lead to significant differences, and can disagree about whether certain band structures can provide benefits. A constant relaxation-time is found to always be optimistic compared to a constant mean-free-path, while the DOS scattering model shows no such clear trend. Notably, the 1D parabolic band and 2D quartic band exhibit the highest power factors with the DOS model, resulting from a rapid decrease in density-of-states, and thus scattering -- suggesting a possible strategy for improved thermoelectrics based on engineering band structures with sharp/discontinuous drops in density-of-states. The DOS scattering approximation also suggests that searches for materials with a delta function-like DOS (as a proxy to the transport distribution) or converged bands may yield limited benefits, due to the increase in scattering. This work highlights the importance of simple and accurate scattering models when rigorous ab-initio scattering calculations are not feasible.  
\end{abstract}

\maketitle

\section{Introduction}
The thermoelectric (TE) conversion efficiency of a material is characterized by its figure-of-merit \cite{Snyder2008}
\begin{align}
	ZT = \frac{S^2 \sigma T}{\kappa_e + \kappa_l}, 
\end{align}
where $S$ is the Seebeck coefficient, $\sigma$ is the electronic conductivity, $\kappa_e$ is the (open-circuit) electronic thermal conductivity, $\kappa_l$ is the lattice thermal conductivity, and $T$ is the temperature. A major focus in thermoelectric research is to identify materials that exhibit especially high $ZT$ values. Theoretical modeling plays an important role through experimental data analysis to help explain observed trends and extract meaningful material parameters \cite{Pettes2013,Kayyalha2016}, and by predicting
promising material candidates to help reduce experimental trial-and-error \cite{Wang2011,Yan2015,Chen2016,Samsonidze2018,Liao2015,Neophytou2017,Jiang2017,Song2017,Hung2017,Zhou2018,Ma2018,Park2019,Askarpour2019}. For bulk materials the TE properties can be calculated using a suitable transport model, most commonly the Boltzmann Transport Equation \cite{Scheidemantel2003,Madsen2006,Kutorasinski2015,Madsen2018,Xing2018} or the Landauer formalism \cite{Kim2009,Jeong2010,Zahid2010,Jeong2012,Maassen2013a,Wick2014,Wick2015}, provided that two quantities are known: the electronic dispersion and scattering rates (here focusing on the {\it electronic} component of TE transport).

Regarding electron band structure, it is common to find both simple analytical models \cite{Minnich2009,Pei2011,Bahk2014,Tang2015,Lin2016} (e.g. effective mass model) and detailed numerical descriptions obtained from first-principles \cite{Scheidemantel2003,Zahid2010,Maassen2013a,Yan2015,Chen2016,Song2017,Xing2018} (e.g. density functional theory (DFT)). The former are particularly convenient when analyzing measured data and useful for providing design strategies \cite{Pei2011,Tang2015}. Moreover, the analytical dispersion models can be straightforwardly benchmarked against more rigorous techniques; in many cases yielding satisfactory agreement (over the energy range where transport occurs, i.e. $\sim$\,10\,$k_B T$). Concerning the scattering rates, rigorous DFT modeling of electron scattering has become more common in the past few years \cite{Giustino2007,Qiu2015,Liao2015,Ponce2016,Zhou2016,Jiang2017,Song2017,Hung2017,Witkoske2017,Zhou2018,Ma2018,Wang2018,Park2019,Askarpour2019}, but these calculations are computationally intensive which hinders their routine usage. Most theoretical investigations rely on one of two simple scattering models: the constant relaxation-time approximation (referred to here as the TAU model) or the constant mean-free-path approximation (MFP model).

Recent first-principles calculations of electron-phonon scattering (the dominant intrinsic scattering mechanism in TE materials) have revealed that the scattering rates often display a trend that closely follows the electron density-of-states (DOS) \cite{Jiang2017,Witkoske2017,Wang2018,Askarpour2019,Pshenay2018,Graziosi2019}, which is not as well captured within a constant scattering-time or mean-free-path assumption \cite{Askarpour2019}. It can be shown that the scattering rates should have the same energy dependence as the DOS in the case of a 3D parabolic band with electron-acoustic phonon scattering (within a deformation potential treatment) \cite{Lundstrom2000}. Rigorous DFT analyses suggest this trend may be more broadly applicable to different materials with complex dispersions. Because DOS calculations are routine and inexpensive, assuming the scattering rates are proportional to the DOS (herein referred to as the DOS model) is an alternative simple scattering model that is more physically routed, and likely more accurate than the commonly used TAU and MFP models.

In this study we investigate how the TE parameters, particularly the power factor $PF=S^2\sigma$, calculated with the DOS model compare to the TAU and MFP models when applied to three example analytical dispersions: parabolic dispersion (in 3D, 2D, 1D), Kane dispersion (in 3D, 2D, 1D), and ring-shaped dispersion (2D). Our findings show that the different scattering models can vary qualitatively in their predictions, which has important consequences when it comes to providing experimental guidance. Moreover, the DOS model points to a potential new strategy for band structure engineering, and suggests that searches for materials with a delta function-like DOS (as a proxy to the transport distribution) or converged bands may not yield the desired benefits, as a result of increased scattering. The paper is organized as follows. In Section \ref{sec:approach} we review our theoretical approach based on the Landauer formalism. Section \ref{sec:results} presents the TE properties of various electronic dispersions using the DOS, TAU, and MFP scattering models. Some discussions are provided in Section \ref{sec:discussion}, before summarizing our findings in Section \ref{sec:conclusions}.

\section{Theoretical approach} 
\label{sec:approach}
In this study the thermoelectric coefficients are calculated, and analyzed, within a Landauer transport framework. The electronic conductivity $\sigma$, Seebeck coefficient $S$, and electronic thermal conductivity $\kappa_e$ are expressed as \cite{Kim2009,Jeong2010}: 
\begin{align}
	\sigma &= \left( \frac{2q^2}{h} \right) I_0, \label{eq:cond} \\  
	S &= -\left( \frac{k_B}{q} \right) \frac{I_1}{I_0}, \label{eq:seebeck} \\ 
	\kappa_e &= \frac{2k_B^2T}{h} \left( I_2 - \frac{I_1^2}{I_0} \right), \label{eq:kappae}
\end{align}
with the quantity $I_j$ defined as
\begin{align}
	I_j = \int_{-\infty}^{\infty}  M(E) \, \lambda(E) \left( \frac{E-\mu}{k_BT}\right)^j \left[-\frac{\partial f_0}{\partial E}\right]\,dE, \label{eq:fermifunc}
\end{align}
where $M(E)$ is the distribution-of-modes (DOM), or number of conducting channels per cross-sectional area, and $\lambda(E)$ is the mean-free-path (MFP) for backscattering, $f_0(E,\mu)$ is the Fermi-Dirac distribution, $\mu$ is the Fermi energy, and $q$ is the magnitude of the electron charge. A Boltzmann equation approach often expresses the TE coefficients in terms of the transport distribution, which is equal to $\Sigma(E)=(2/h) M(E)\lambda(E)$ \cite{Jeong2010,Witkoske2017} and closely related to the integrand in Eq.~(\ref{eq:fermifunc}). While the physical meaning of $\Sigma(E)$ is not so transparent, both $M(E)$ and $\lambda(E)$ have a clear physical interpretation, as discussed below.

All material specific properties are captured in $M(E)$ and $\lambda(E)$. The distribution-of-modes is a measure of a material's intrinsic ability to carry current, and closely related to the ballistic conductance. The mean-free-path for backscattering is defined as the average distance travelled along the transport direction, here taken as $x$, before scattering changes the sign of the $p_x$ component of the carrier momentum. $M(E)$ and $\lambda(E)$ are defined as \cite{Jeong2010,Lundstrom2013}:
\begin{align}
  M(E) &= \frac{h}{4} \sum_{k,s} |v_x(k)|\,\delta(E-\epsilon(k)) \,/\, \Omega, \label{eq:modes} \\
  \lambda(E) &= 2 \, \frac{\sum_{k,s} v_x^2(k)\,\tau(k)\,\delta(E-\epsilon(k))}{\sum_{k,s} |v_x(k)|\,\delta(E-\epsilon(k))}, \label{eq:mfp} 
\end{align}
where $\tau(k)$ is the scattering time of state $k$ in the Brillouin zone, $v_x = (1/\hbar) \partial\epsilon(k)/\partial k_x$ is the group velocity, and $\epsilon(k)$ is the electron dispersion (sums are over reciprocal lattice points $k$ in the Brillouin zone and spin $s$). In order to calculate $M(E)$ and $\lambda(E)$, and hence all the TE coefficients, two pieces of information are required: the electron dispersion and the scattering physics (captured in $\tau$).

While band structure models are commonly available or readily calculated -- both analytical and numerically computed -- it is generally more challenging to accurately capture the scattering time. Rigorously resolving the relaxation time $\tau(k)$ over the entire Brillouin zone is possible \cite{Giustino2007,Qiu2015,Liao2015,Ponce2016,Zhou2016,Jiang2017,Song2017,Hung2017,Witkoske2017,Zhou2018,Ma2018,Wang2018,Park2019,Askarpour2019}, but computationally expensive, thus various approximations are commonly adopted. One of the most widespread approximations involves setting the scattering time to a constant, $\tau(k)=\tau_0$ (TAU model). Another common approximation is to set the mean-free-path to a constant, $\lambda(E)=\lambda_0$ (MFP model). [We note that $\lambda$ differs from the standard definition of mean-free-path, $l(k)=|v(k)|\tau(k)$ -- for isotropic bands they are related by a numerical factor, $\lambda=(4/3) l$ \cite{Jeong2010}]. Generally, the TAU and MFP models are adopted when the TE properties are calculated within the Boltzmann and Landauer transport frameworks, respectively, where they are most conveniently implemented. We note that TE parameters calculated with both the Boltzmann and Landauer approaches yield identical results \cite{Jeong2010}, when the relaxation time approximation is adopted. A constant $\lambda$ can be derived assuming a 3D parabolic band and a deformation potential treatment of acoustic phonon scattering \cite{Lundstrom2000}, however a constant $\tau$ is more difficult to physically justify.

Phonon and impurity scattering are typically the dominant collision mechanisms for electrons in thermoelectrics. Rigorous first-principles calculations of electron-phonon scattering have shown that often the scattering rates are (roughly) proportional to the electron DOS, $D(E)$ \cite{Jiang2017,Witkoske2017,Wang2018,Pshenay2018,Askarpour2019}. Intuitively, one would expect the probability of scattering to scale with the number of available final states, which partially explains this trend. Non-polar phonon scattering can lead to scattering with an energy dependence similar to the electron DOS \cite{Fischetti1991}, however polar phonon and charged impurity scattering can have trends that are opposite to the DOS \cite{Lundstrom2000}. When there is a relatively large density of mobile carriers, as is the case in most optimized thermoelectric materials, screening (e.g. within Thomas-Fermi theory) can play an important role and alter the scattering characteristics resulting in an energy dependence in agreement with the DOS \cite{Lundstrom2000,Askarpour2019}. For these reasons, we believe that the DOS scattering model will more often better approximate the true scattering properties in optimized thermoelectrics versus the MFP and TAU models. This, however, is not a general rule and may not hold in lightly-doped semiconductors or for different scattering mechanisms not considered here.

\begin{figure*}	
	\includegraphics[width=17cm]{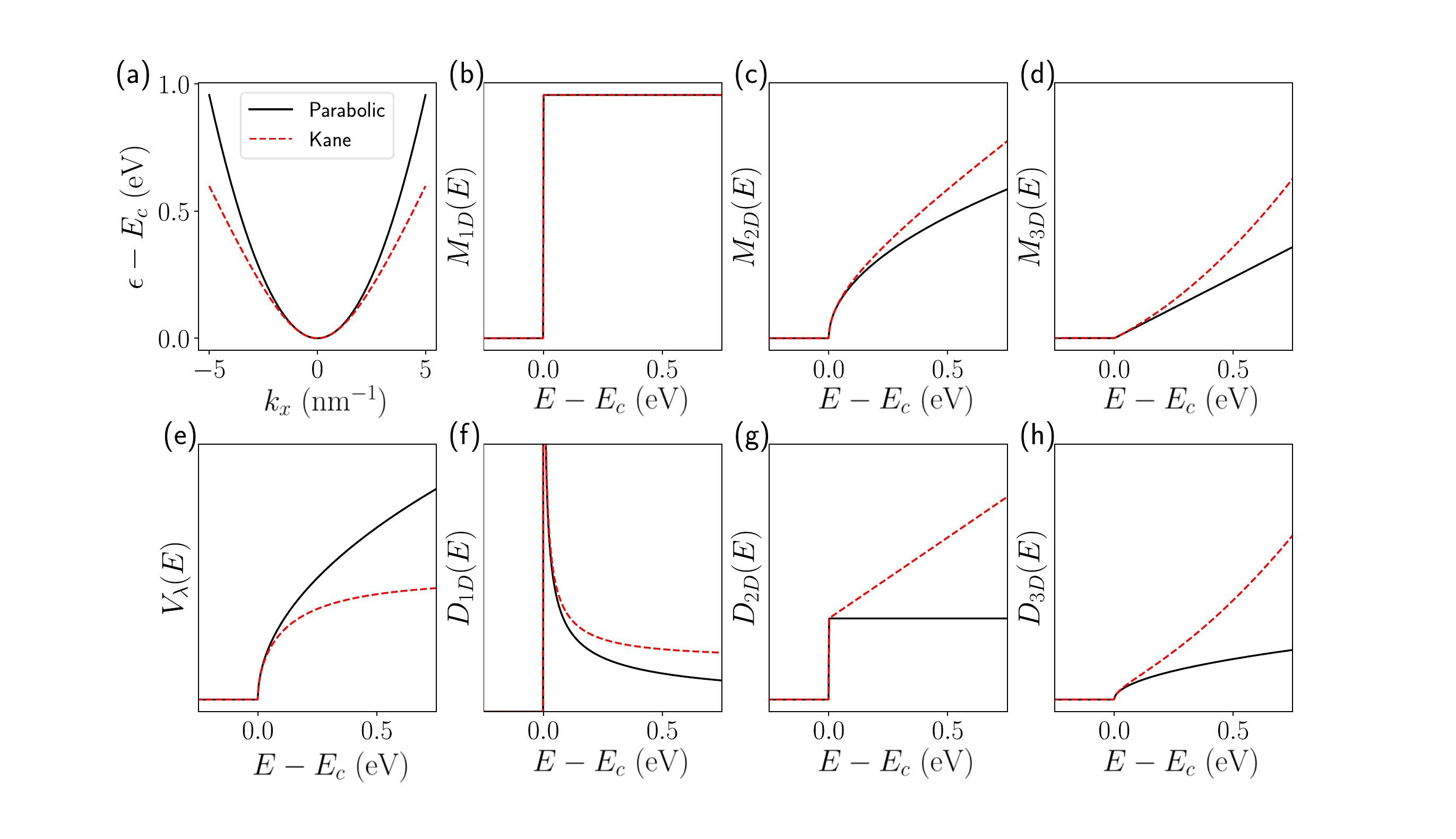}
	\caption{Parabolic and Kane bands in 1D, 2D and 3D ($m^*$\,=\,$m_0$, $\alpha$\,=\,1 eV$^{-1}$). Sketches of (a) electron dispersion $E(k)$, (b,c,d) distribution-of-modes $M(E)$, (f,g,h) density-of-states $D(E)$, and (e) average velocity $V_{\lambda}(E)$ (see main text for definition). Parabolic band and Kane band results appear as black solid and dashed red lines, respectively.} \label{fig:para_kane}
\end{figure*}

\begin{table*}[t]
	\begin{tabular}{|c || c | c  |c | c | c |c |} 
		\hline
		Dimension/Quantity & $M(E)$ & $V_\lambda(E)$ & $D(E)$ & $\Sigma_{\text{MFP}}(E)$ &$\Sigma_{\text{TAU}}(E)$ &$\Sigma_{\text{DOS}}(E)$  \\ [0.5ex] 
		\hline\hline
		1D & $1$ & $\sqrt{E}$ & 1/$\sqrt{E}$ & $1$ & $\sqrt{E}$ & $E$ \\ 
		\hline
		2D & $\sqrt{E}$ & $\sqrt{E}$ & $1$ & $\sqrt{E}$ & $E$ & $E$ \\
		\hline
		3D & $E$ & $\sqrt{E}$ & $\sqrt{E}$ & $E$ & $E^{3/2}$ & $E$\\
		\hline
	\end{tabular}
	\caption{Isotropic parabolic band in 1D, 2D and 3D. Energy dependence of distribution-of-modes $M(E)$, average velocity $V_{\lambda}(E)$, density-of-states $D(E)$, and transport distribution $\Sigma(E)$ (within the MFP, TAU and DOS scattering models). For full expressions, refer to Appendix~\ref{app:para}.}
	\label{tab:para}
\end{table*}

The DOS scattering approximation, wherein the scattering rates take the form $1/\tau(E)$\,=\,$K_0\,D(E)$ \cite{Witkoske2017}, and its connection to physical scattering expressions has been highlighted in previous studies by, for example, Allen et al. \cite{Allen1976,Allen1986} and Fischetti \cite{Fischetti1991}. Recent studies in which DOS scattering has been adopted include \cite{Zhou2011,Kumarasinghe2019,McKinney2017,Putatunda2019}, but are far from being as common as the MFP or TAU approximations. There have also been investigations comparing the DOS scattering model to other simple scattering models \cite{Witkoske2017,Kumarasinghe2019,Zhou2011,Putatunda2019}.

For accurate predictions of TE properties a detailed consideration of scattering is required. Often both phonon and ionized impurity scattering are simultaneously present, in which case there can be an interplay where either type can be dominant at certain electron energies. With independent scattering processes, the total collision rates are obtained by summing the rates of each mechanism, following Matthiessen's rule. Since charged impurity and polar phonon scattering can have large scattering rates near the band edge \cite{Lundstrom2000}, this may not be well captured by the DOS model which in 3D materials tends to zero at the band edge (strong screening, however, can significantly reduce these rates near the band edge). This interplay between different mechanisms can be optimized to enhance TE characteristics. For example, ionized impurity scattering can increase the Seebeck coefficient and power factor, as a result of its particular energy dependence aided also by small-angle scattering, although if the magnitude of the scattering is too large carrier mobility will be deteriorated. Reducing ionized impurity scattering, such that it acts in combination with phonon scattering, can lead to increases in mobility and power factor, as demonstrated in \cite{Shuai2017, Mao2017}.

In this work, the three aforementioned simple scattering models (MFP, TAU and DOS scattering models) will be compared -- with the DOS model believed to be the most physical option -- in the case of a few simple analytical band structures. The transport distribution, while not having a straightforward physical interpretation, is a useful quantity for understanding which materials should provide good TE performance (this will be discussed later). The equations below show the different transport distributions that arise when making use of the MFP, TAU and DOS scattering model approximations:
\begin{align}
	\Sigma_{\text{MFP}}(E) &= \frac{2}{h}M(E) \cdot \lambda_0 , \label{eq:td_mfp} \\
	\Sigma_{\text{TAU}}(E) &= \frac{2}{h}M(E) \cdot V_\lambda(E) \cdot \tau_0 , \label{eq:td_tau} \\
	\Sigma_{\text{DOS}}(E) &= \frac{2}{h}M(E) \cdot V_\lambda(E) \cdot \frac{1}{K_0\,D(E)} , \label{eq:td_dos}
\end{align}
where
\begin{align}
 V_{\lambda}(E) = 2 \,\frac{\sum_{k,s}v_x^2(k) \,\delta(E-\epsilon(k))}{\sum_{k,s}|v_x(k)|\,\delta(E-\epsilon(k))},
\end{align}
is an averaged velocity (defined such that $\lambda(E)=V_{\lambda}(E) \tau(E)$), and $\lambda_0$, $\tau_0$ and $K_0$ are constants. Note that only three distinct energy-dependent quantities appear in the above expressions: $M(E)$, $D(E)$, and $V_\lambda(E)$. Figure \ref{fig:para_kane} and Table \ref{tab:para} both display the energy-dependence of these quantities in the case of a single isotropic parabolic band.

$\lambda_0$, $\tau_0$ and $K_0$ can be treated as free parameters in each model. Typically the value of the parameter, depending on which model is adopted, is chosen to reproduce experimental data, for example the electronic conductivity (for a given Fermi or doping level). We opt to compare the three scattering models by having them agree about the average mean-free-path for backscattering when the Fermi level is at the conduction band edge, $\mu$\,=\,$E_C$ \cite{Jeong2010}:
\begin{align}
	\langle \langle \lambda\rangle \rangle _{\mu=E_C} = \frac{\int_{-\infty}^{\infty} M(E) \lambda(E) \left[-\frac{\partial f_0}{\partial E}\right] dE}{\int_{-\infty}^{\infty} M(E) \left[-\frac{\partial f_0}{\partial E}\right] dE}  = \lambda_0. \label{eq:av_mfp} 
\end{align}
This is equivalent to fixing the electronic conductivity at the band edge. In this work we set $\langle\langle\lambda\rangle\rangle$ to 10 nm (a typical value), and take our zero energy reference to be the band edge, $E_C$.

\section{Thermoelectric Properties}
\label{sec:results}
The analysis that follows can be thought of as answering the following question: To what extent do the predictions of the MFP and TAU models differ from those of the more physical DOS model, when made to agree about the electrical conductivity (when $\mu$\,=\,$E_C$)? To explore this question we have chosen three example electron dispersions, including a parabolic band (in 3D, 2D, 1D), Kane band (in 3D, 2D, 1D) and ring-shaped band (2D) -- this list is not exhaustive but selected to illustrate how the different scattering models behave. We begin with a parabolic (effective mass) band, which will serve as a starting point and reference for our analysis. After, we will examine how deviations from parabolicity, captured with the Kane model, influence the scattering and TE properties. Lastly, a fundamentally different type of dispersion, a ring-shaped or ``mexican-hat'' band model, will be explored. Our analysis will focus mainly on how the scattering models influence the power factor $PF=S^2\sigma$, a key TE metric that depends solely on electronic transport.

\subsection{Parabolic (effective mass) dispersion}
\label{sec:parabolic} 
\begin{figure}	
	\includegraphics[width=8.5cm]{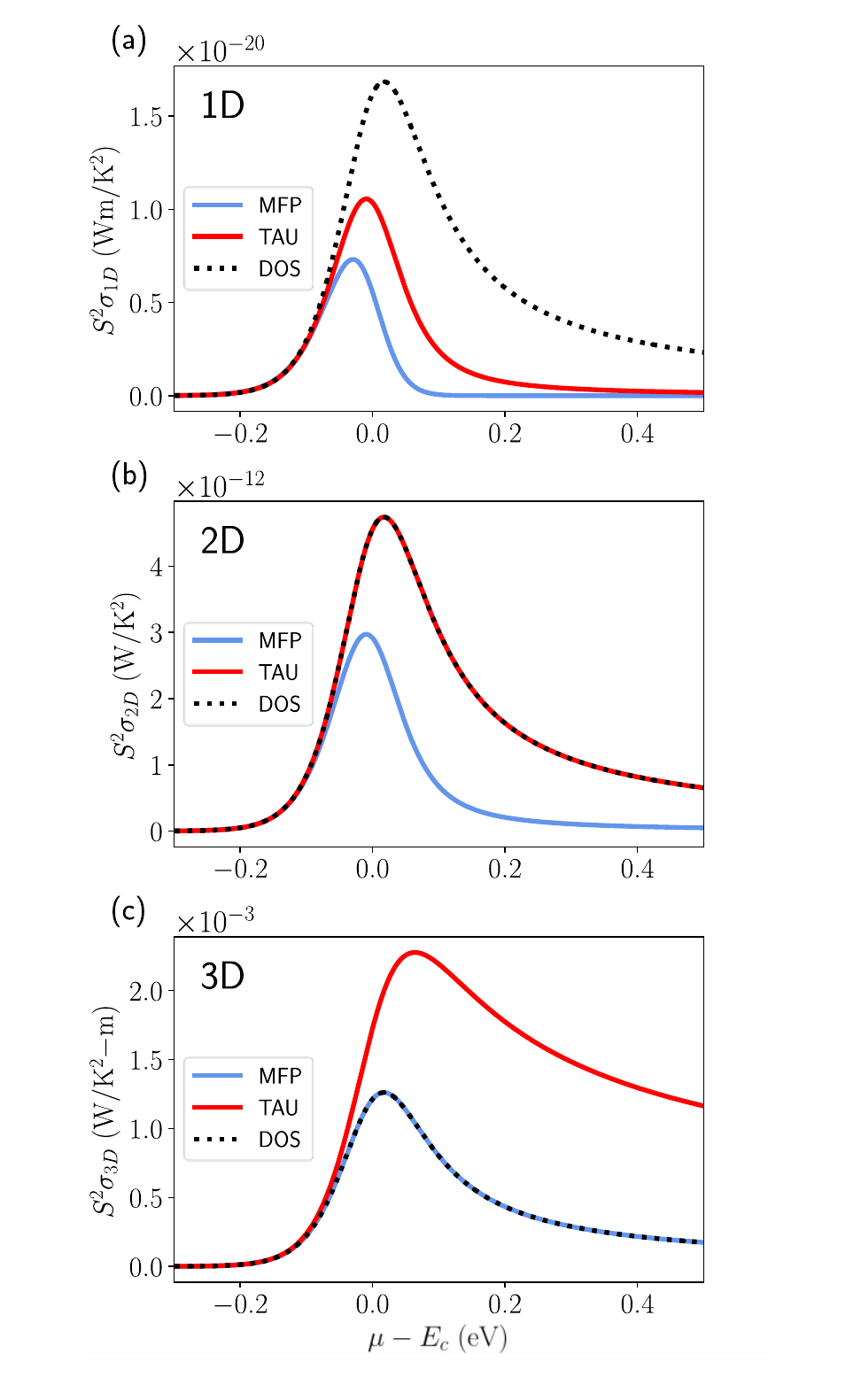}
	\caption{Isotropic parabolic band. Power factor $PF$ versus Fermi level $\mu$ in (a) 1D, (b) 2D, and (c) 3D, using the MFP, TAU and DOS scattering models. For these calculations, $m^*$\,=\,$m_0$, $T$\,=\,300~K, and $\langle\langle\lambda\rangle\rangle_{\mu=E_C}$\,=\,10~nm.}  \label{fig:para_pf}
\end{figure}

We begin with perhaps the most familiar electron dispersion model -- a single isotropic parabolic, effective mass band (see for example Refs.~\cite{Minnich2009,Pei2011,Bahk2014,Tang2015,Lin2016,Witkoske2017}). The electron dispersion has the form:
\begin{align}
\epsilon(k)=\frac{\hbar^2 k^2}{2m^*}, \label{eq:para_ek}
\end{align}
where $m^*$ is the effective mass. Using Eqns.~(\ref{eq:cond})-(\ref{eq:mfp}), we calculate the TE properties. Figure~\ref{fig:para_pf} presents the power factors of a parabolic band with $m^*$\,=\,$m_0$ (free electron mass), calculated using all three scattering approximations in 1D, 2D and 3D. The results for the MFP model were previously shown in Ref.~\cite{Kim2009}. Note that the units of $PF$ and $\sigma$ vary across dimensions -- a direct comparison requires introducing an ``effective cross-sectional area'' for the low-dimensional materials. This effective area is a function of how densely the low-dimensional materials can be packed without distorting the electronic dispersion, which is highly material-dependent, so we do not attempt to define one for the models considered in this work.

\begin{figure}	
	\includegraphics[width=8.5cm]{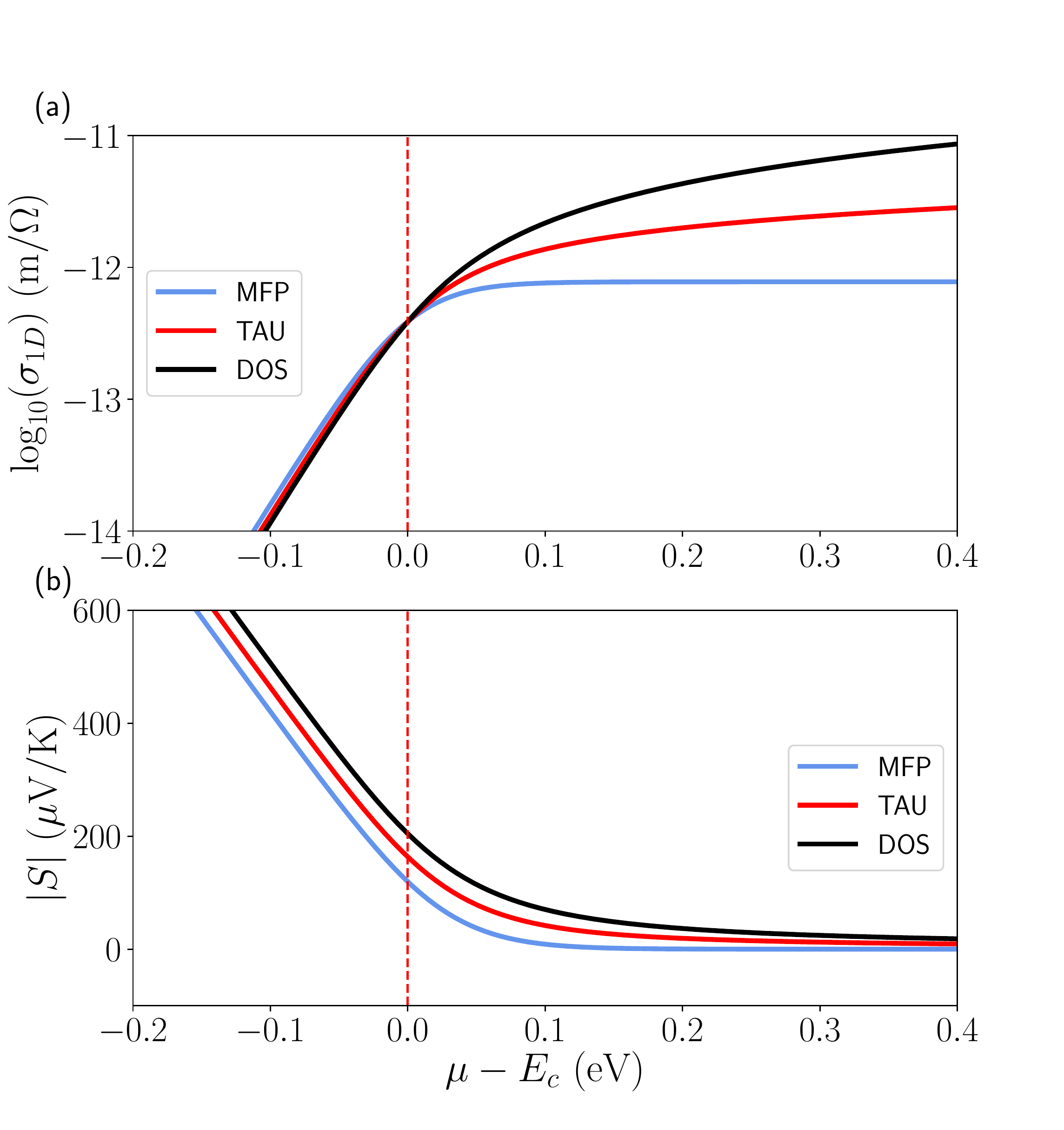}
	\caption{1D isotropic parabolic band. (a) Electrical conductivity and (b) Seebeck coefficient versus Fermi level $\mu$, using the MFP, TAU and DOS scattering models. For these calculations, $m^*$\,=\,$m_0$, $T$\,=\,300~K, and $\langle\langle\lambda\rangle\rangle_{\mu=E_C}$\,=\,10~nm.}  \label{fig:1D_seebeck}
\end{figure}

We focus first on the one-dimensional case. It is noteworthy that despite being made to agree about the electronic conductivity (when $\mu=E_C$), the DOS model predicts a power factor more than 50\% higher than that of the TAU model, which in turn is larger than that of the MFP model. This increased performance comes from the fact that the DOS for a 1D parabolic band is a decreasing function with respect to energy ($1/\sqrt{E}$). This results in the DOS model predicting a lower scattering rate for high-energy electrons, which increases the electrical conductivity and Seebeck coefficient when $\mu>E_C$, as shown in Fig.~\ref{fig:1D_seebeck}. The differences among the scattering models arise from the mean-free-paths for backscattering: $\lambda_{\rm DOS}^{\rm 1D}(E)\propto E^1$, $\lambda_{\rm TAU}^{\rm 1D}(E)\propto E^{1/2}$, and $\lambda_{\rm MFP}^{\rm 1D}(E)\propto E^0$. As the Fermi level moves deeper in the band, the larger magnitude of $\lambda(E)$ increases $\sigma$, and the larger energy dependence of $\lambda(E)$ increases the average energy at which current flows which is proportional to $S$ \cite{Lundstrom2013} (this point is discussed below in terms of the transport distribution).

For the two-dimensional case in Fig.~\ref{fig:para_pf}, the predictions of the DOS model and the TAU model are identical. This is because the DOS of a 2D parabolic band is constant, which results in a constant scattering time in the DOS model. The MFP model is again the most pessimistic of the three. In the three-dimensional case, it is now the MFP model that agrees with the DOS model. This occurs because in 3D, $V_\lambda(E)$ and $D(E)$ have the same energy dependence, namely $\sqrt{E}$, so their contributions to $\lambda_{\text{DOS}}(E) = V_\lambda(E)/(K_0\,D(E))$ cancel out leaving a constant.

When comparing the magnitude of the power factors to the energy dependence of the corresponding transport distributions (shown in Table~\ref{tab:para}), it is clear that both are correlated: a $\Sigma(E)$ with stronger energy dependence results in a larger maximum $PF$. To illustrate why this happens, we introduce the Fermi window function. From Eqns.~(\ref{eq:cond})-(\ref{eq:seebeck}), we see that $\sigma\propto I_0$ and $S\propto I_1$, thus it is convenient to define the following function:
\begin{align}
	W_j(E,\mu) = \left(\frac{E-\mu}{k_B T}\right)^j \left[-\frac{\partial f_0(E, \mu)}{\partial E}\right], \label{eq:fermiwindow}
\end{align}
referred to as the Fermi window function of $j^{\rm th}$ order, which appears in the integral of $I_j$ (Eq.~(\ref{eq:fermifunc})). One can show that $\int_{-\infty}^{\infty} W_0\,dE=1$ and $\int_{-\infty}^{\infty} W_1\,dE=0$. Since the Fermi window functions are not material specific, the TE quantities are determined by the transport distribution, $\Sigma(E) = (2/h)M(E)\lambda(E)$.

\begin{figure}	
\includegraphics[width=8.5cm]{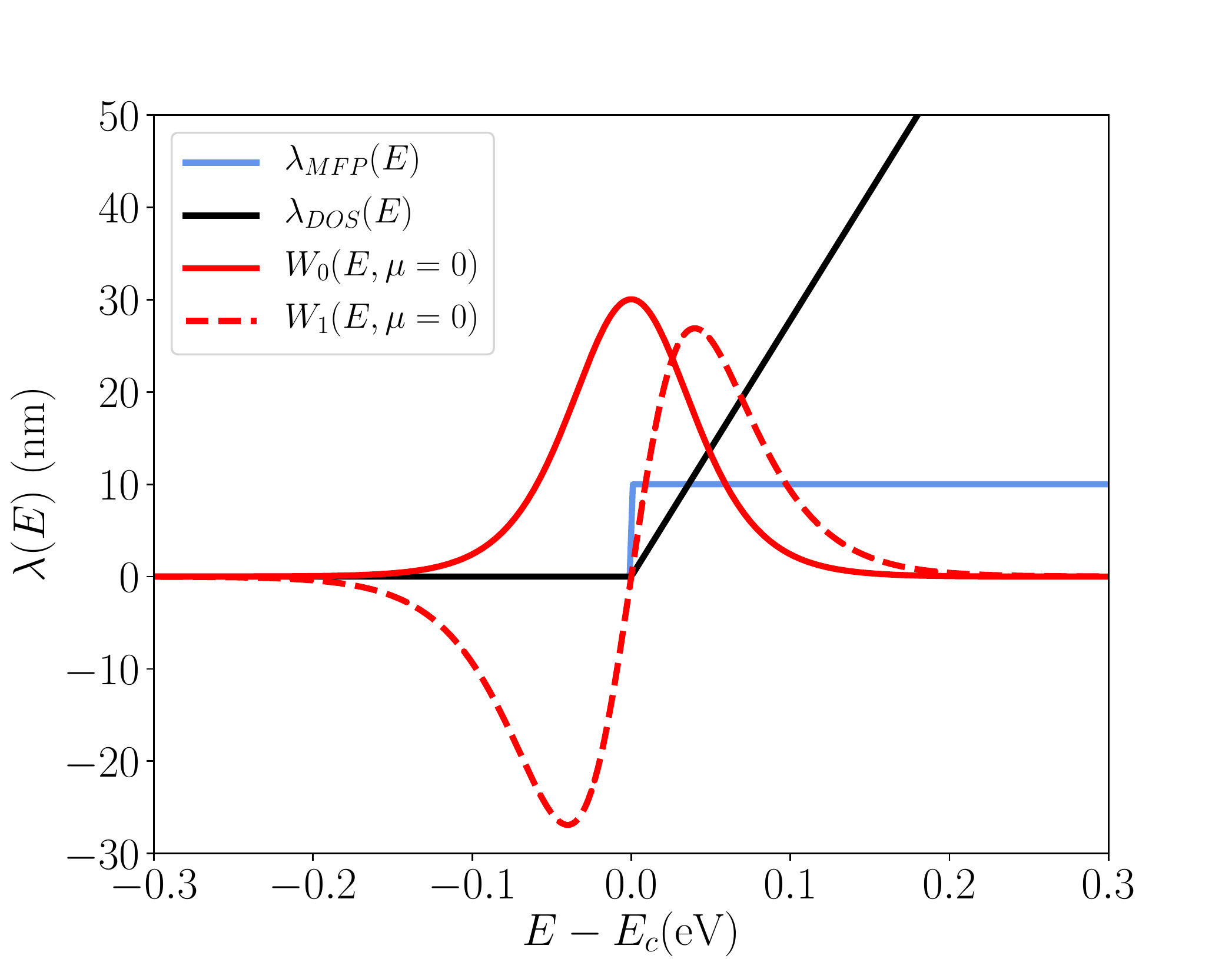}
\caption{1D isotropic parabolic band. Mean-free-path for backscattering versus energy for the MFP and DOS models. Also shown are the 0$^{\rm th}$- and 1$^{\rm st}$-order Fermi windows $W_{0,1}$ (not to scale).  For these calculations, $\mu$\,=\,$E_C$, $m^*$\,=\,$m_0$, $T$\,=\,300~K, and $\langle\langle\lambda\rangle\rangle$\,=\,10~nm.} \label{fig:mfp_fermifunc}
\end{figure}

Figure~\ref{fig:mfp_fermifunc} presents $\lambda(E)$ versus energy for the MFP and DOS scattering models, for a 1D parabolic band. Since the DOM in 1D is a constant, the mean-free-path for backscattering is a proxy for the transport distribution, $\Sigma(E)\propto M(E) \lambda(E) \propto \lambda(E)$ -- more generally when the DOM is not a constant, the following discussion holds when analyzing $\Sigma(E)$ instead of $\lambda(E)$. Also plotted are the 0$^{\rm th}$- and 1$^{\rm st}$-order Fermi windows, $W_{0,1}(E)$ (not to scale). The average mean-free-path for backscattering, related to the integral of $\lambda(E) W_0(E)$, is the same for both models and sets the magnitude of $\lambda(E)$ (and $\Sigma(E)$). The Seebeck coefficient, however, depends on the integral of $\lambda(E) W_1(E)$, where $W_1(E)$ favors a strong asymmetry above and below the Fermi level. Even when all conduction occurs above the Fermi level, as is the case in Fig.~\ref{fig:mfp_fermifunc}, because $W_1(E)$ extends further in energy compared to $W_0(E)$, the $\lambda(E)$ (or $\Sigma(E)$) with larger energy dependence will yield a larger Seebeck coefficient. This explains the observed trend in Fig.~\ref{fig:para_pf}. From this perspective, we find that the transport distribution is a useful quantity for assessing the potential of a given material, where its magnitude controls the electrical conductivity and the degree of asymmetry (energy-dependence) controls the Seebeck coefficient.

\begin{table*}
	\begin{tabular}{|c || c | c  |c | c | c |c |} 
		\hline
		Dimension/Quantity & $M(E)$ & $V_\lambda (E)$ & $D(E)$ & $\Sigma_{\text{MFP}}(E)$ &$\Sigma_{\text{TAU}}(E)$ &$\Sigma_{\text{DOS}}(E)$  \\ [0.5ex] 
		\hline\hline
		1D & $1$ & $\frac{\sqrt{E(1+\alpha E)}}{1+ 2\alpha E}$ & $\frac{1+ 2\alpha E}{\sqrt{E(1+\alpha E)}}$ & $1$ & $\frac{\sqrt{E(1+\alpha E)}}{1+ 2\alpha E}$ & $\frac{E(1+\alpha E)}{(1+ 2\alpha E)^2}$ \\ 
		\hline
		2D & $\sqrt{E(1+\alpha E)}$ & $\frac{\sqrt{E(1+\alpha E)}}{1+ 2\alpha E}$ & $1+ 2\alpha E$ & $\sqrt{E(1+\alpha E)}$ & $\frac{E(1+\alpha E)}{1+ 2\alpha E}$ & $\frac{E(1+\alpha E)}{(1+ 2\alpha E)^2}$ \\
		\hline
		3D & $E(1+\alpha E)$ & $\frac{\sqrt{E(1+\alpha E)}}{1+ 2\alpha E}$ & $\sqrt{E(1+\alpha E)}(1+ 2\alpha E)$ & $E(1+\alpha E)$ & $\frac{[E(1+\alpha E)]^{3/2}}{1+ 2\alpha E}$ & $\frac{E(1+\alpha E)}{(1+ 2\alpha E)^2}$\\
		\hline
	\end{tabular}
	\caption{Kane band in 1D, 2D and 3D. Energy dependence of distribution-of-modes $M(E)$, average velocity $V_{\lambda}(E)$, density-of-states $D(E)$, and transport distribution $\Sigma(E)$ (within the MFP, TAU and DOS scattering models). For full expressions, refer to Appendix~\ref{app:kane}.}
	\label{tab:kane}
\end{table*}

From this observation follows a noteworthy corollary: when made to agree about $\langle\langle\lambda\rangle\rangle$ in a semiconducting material, \textit{a TAU scattering model will be more optimistic than a MFP scattering model}. While $\Sigma_{\rm MFP}(E)$ is simply proportional to $M(E)$, $\Sigma_{\rm TAU}(E)$ is proportional to $M(E) V_{\lambda}(E)$ (see Eqns.~(\ref{eq:td_mfp})-(\ref{eq:td_tau})). The latter will \textit{always} go to zero at the band edge, since the band edge corresponds to a local minimum in the electronic dispersion in $k$-space, and hence has vanishing velocity. As such, $V_{\lambda}(E)$ will always be an increasing quantity, at least for low energies. (This may not be the case for metals, or materials with linear bands such as graphene or topological insulators, which have been proposed as good thermoelectrics \cite{Markov2018}.) This means that, with semiconductors, $\Sigma_{\rm TAU}(E)$ is generally expected to exhibit a larger energy-dependence than $\Sigma_{\rm MFP}(E)$, and hence predict better TE performance.

This observation has important consequences when comparing the predictions based on the constant relaxation-time approximation, often adopted with a Boltzmann approach, versus those based on the a constant mean-free-path approximation, commonly adopted with a Landauer approach. Even when in complete agreement about the electronic dispersion and electrical conductivity of a particular material, the TAU scattering model will inevitably draw more optimistic conclusions than the MFP model. No such general trend exists when comparing to a DOS scattering model, which can be either optimistic or pessimistic depending on the details of the dispersion, as shown for a parabolic band and to be confirmed with other dispersions below.

Our expressions for isotropic bands can be generalized to describe anisotropic parabolic bands as well, as shown in Appendix~\ref{app:para}. It is interesting to note that in the anisotropic case, the transport distributions of the three scattering models each have a different dependence on the effective mass components: $\Sigma_{\rm MFP} \propto \sqrt{m_y m_z}$, $\Sigma_{\rm TAU} \propto \sqrt{m_y m_z/m_x}$, $\Sigma_{\rm DOS} \propto 1/m_x$ (with transport assumed along $x$). This illustrates how the choice in scattering model can be important -- for example, the MFP model tells us that increasing the transverse effective masses $m_{y,z}$ is beneficial, while according to the DOS model only decreasing the longitudinal $m_x$ can have a positive effect. Lastly, in this work we have focused on a single band, however good thermoelectrics can contain multiple bands, or bands with high degeneracy. Implicit in the assumption of DOS scattering is that the coupling constant for intra-valley and inter-valley scattering processes are equal \cite{Witkoske2017,McKinney2017,Kumarasinghe2019}, which can lead to significant differences with the TAU and MFP models -- for example Ref.~\cite{Kumarasinghe2019} compared these scattering models in the case of multiple bands. In general, band convergence strategies are likely to be more pessimistic within the DOS scattering approximation, versus the MFP and TAU models, since the enhanced DOS will result in additional scattering.

\subsection{Kane dispersion}
\label{sec:kane}
Next, we set out to investigate what each scattering model concludes about the effect of deviations from parabolicity. A widely applicable generalization of the parabolic band model is the Kane model, in which the electronic dispersion is modelled as
\begin{align}
	\epsilon(1+\alpha \epsilon) = \frac{\hbar^2 k^2}{2m^*}, \label{eq:kane_ek}
\end{align}
where in addition to the effective mass $m^*$ there is the non-parabolicity parameter $\alpha$. As illustrated in Fig.~\ref{fig:para_kane}, the Kane model gives an electronic dispersion that is parabolic near the band edge, but approaches linearity at higher energies -- $\alpha$ is a measure of how ``linearized'' the band is. This simple model can be derived from $k\cdot p$ theory, and in many cases is found to represent actual electronic band structures more accurately than a pure parabolic band model \cite{Maassen2013b}.

\begin{figure}
	\includegraphics[width=8.5cm]{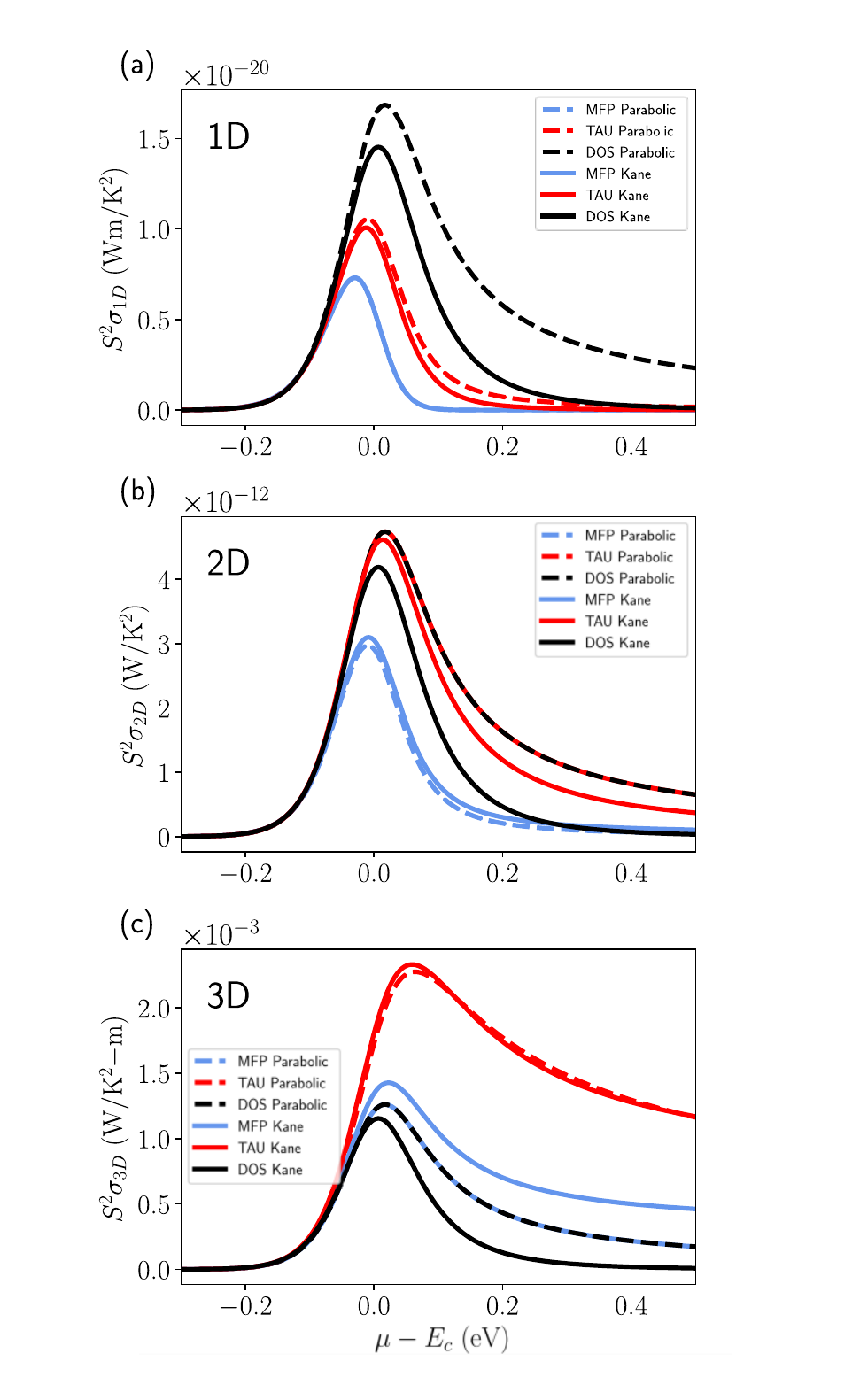}
	\caption{Kane band. Power factor $PF$ versus Fermi level $\mu$ in (a) 1D, (b) 2D, and (c) 3D, using the MFP, TAU and DOS scattering models. The results of an isotropic parabolic band are presented as dashed lines. For these calculations, $m^*$\,=\,$m_0$, $\alpha$\,=\,1.0 eV$^{-1}$, $T$\,=\,300~K, and $\langle\langle\lambda\rangle\rangle_{\mu=E_C}$\,=\,10~nm.} \label{fig:kane_pf}
\end{figure}

Figure~\ref{fig:kane_pf} compares the power factors of a Kane dispersion (in 1D, 2D and 3D) for the three scattering models. The parabolic band results (dashed lines) are also plotted to serve as a reference ($m^*$\,=\,$m_0$, and $\alpha$\,=\,1.0 eV$^{-1}$ for the Kane band, which is a typical value of $\alpha$ \cite{Maassen2013b}). The changes in $PF$ can be understood in terms of the effect that the ``linearization'' of the dispersion has on $M(E)$, $D(E)$, and $V_{\lambda}(E)$, which are shown as dashed lines in Fig.~\ref{fig:para_kane}.

Firstly, the MFP model is consistently optimistic about the effect of ``linearizing'' a parabolic band, meaning the $PF$ increases with the Kane parameter $\alpha$. This is because, at a given energy, a Kane band will have a larger $M(E)$ than a parabolic band with the same effective mass. (One exception is the 1D case, where $M(E)$ is a step function for both parabolic and Kane bands.) However, the TAU model is pessimistic in 1D and 2D, and optimistic in 3D. This difference comes from the fact that a Kane band is slower than its parabolic band equivalent (i.e. same $m^*$), thus reducing $V_{\lambda}(E)$ -- a fact that the MFP model is blind to. In 3D, $M(E)$ increases by enough to offset the detrimental effects of lower velocity electrons, but in lower dimensions this is no longer the case.

DOS scattering, however, is consistently pessimistic about the effects of linearization. This occurs since, in addition to the lower velocity, the increased DOS results in more scattering -- something that the other two scattering models are blind to. This combination of slower states and increased scattering is enough to outweigh the performance benefits of an increased number of conducting channels, regardless of spatial dimension.

While none of the scattering models predict especially favorable TE performance from a Kane band, it is nonetheless noteworthy that the simplified MFP and TAU model approximations can occasionally draw the exact opposite conclusion of the more physical DOS scattering model, namely that Kane-like deviations from parabolicity can result in improved performance. DOS scattering is consistent across all dimensions in its prediction that a Kane band will perform worse than a parabolic band. Lastly, we note that the conclusions of this section were drawn from comparing a parabolic band to a Kane band with a representative value of $\alpha$, which may not be the same for other arbitrary values of $\alpha$.

\begin{table*}
	\begin{tabular}{|c || c | c  |c | c | c |c |} 
		\hline
		Energy Range & $M(E)$ & $V_\lambda(E)$ & $D(E)$ & $\Sigma_{\rm MFP}(E)$ &$\Sigma_{\rm TAU}(E)$ &$\Sigma_{\rm DOS}(E)$  \\ [0.5ex] 
		\hline\hline
		$E < \epsilon_0$ & $\sqrt{1 + \sqrt{\frac{E}{\epsilon_0}}}+\sqrt{1 - \sqrt{\frac{E}{\epsilon_0}}}$ & $\frac{2 \sqrt{E}}{\sqrt{1 + \sqrt{\frac{E}{\epsilon_0}}}+\sqrt{1 - \sqrt{\frac{E}{\epsilon_0}}}}$ & $2\sqrt{\frac{\epsilon_0}{E}}$ & $\sqrt{1 + \sqrt{\frac{E}{\epsilon_0}}}+\sqrt{1 - \sqrt{\frac{E}{\epsilon_0}}}$ & $2\sqrt{E}$ & $E$ \\ 
		\hline
		$E > \epsilon_0$ & $\sqrt{1 + \sqrt{\frac{E}{\epsilon_0}}}$ & $\sqrt{E} \sqrt{1 + \sqrt{\frac{E}{\epsilon_0}}}$ & $\sqrt{\frac{\epsilon_0}{E}}$ & $\sqrt{1 + \sqrt{\frac{E}{\epsilon_0}}}$ & $\sqrt{E}(1+\sqrt{\frac{E}{\epsilon_0}})$ & $E(1+\sqrt{\frac{E}{\epsilon_0}})$ \\
		\hline
	\end{tabular}
    \caption{Quartic band. Energy dependence of distribution-of-modes $M(E)$, average velocity $V_{\lambda}(E)$, density-of-states $D(E)$, and transport distribution $\Sigma(E)$ (within the MFP, TAU and DOS scattering models). When present, the factors of 2 serve to enforce the continuity (or lack thereof) of the distributions. For full expressions, refer to Appendix~\ref{app:quartic}.}
\end{table*}

\subsection{Ring-shaped dispersions}
\label{sec:ringshaped}
Next, we consider a more exotic class of 2D band structures, sometimes referred to as ``ring-shaped'', ``warped'' or ``Mexican-hat'' bands \cite{Wick2015}, that arise in few-layer 2D materials. This type of $\epsilon(k)$ is qualitatively different from more common parabolic/Kane dispersions, in that the band edge doesn't correspond to a {\it point} in $k$-space but rather a {\it line} (a ring) in $k$-space, which gives rise to distinct properties. Previous studies have proposed that the TE characteristics of such ring-shaped band materials (e.g. monolayer Bi, bilayer graphene, few-layer Bi$_2$Te$_3$) would outperform those of standard dispersions \cite{Zahid2010,Maassen2013a,Wick2015} -- with the benefits coming from a rapid, discrete increase in the DOM at the band edge. However, the previous analyses relied on either the MFP or TAU scattering models, which we have shown can differ significantly from the more physical DOS model. Here, we revisit the performance of this type of dispersion by comparing all three scattering models, using approximate analytical descriptions for the $\epsilon(k)$. We also note that beyond the ring-shaped dispersions considered here, other non-simple band structures have also been proposed to be beneficial for TE performance, including ``pudding-mold'' \cite{Usui2013,Mori2013}, ``camel-back'' \cite{Wang2014} and non-parabolic \cite{Chen2013} dispersions.

\begin{figure}	
	\includegraphics[width=7cm]{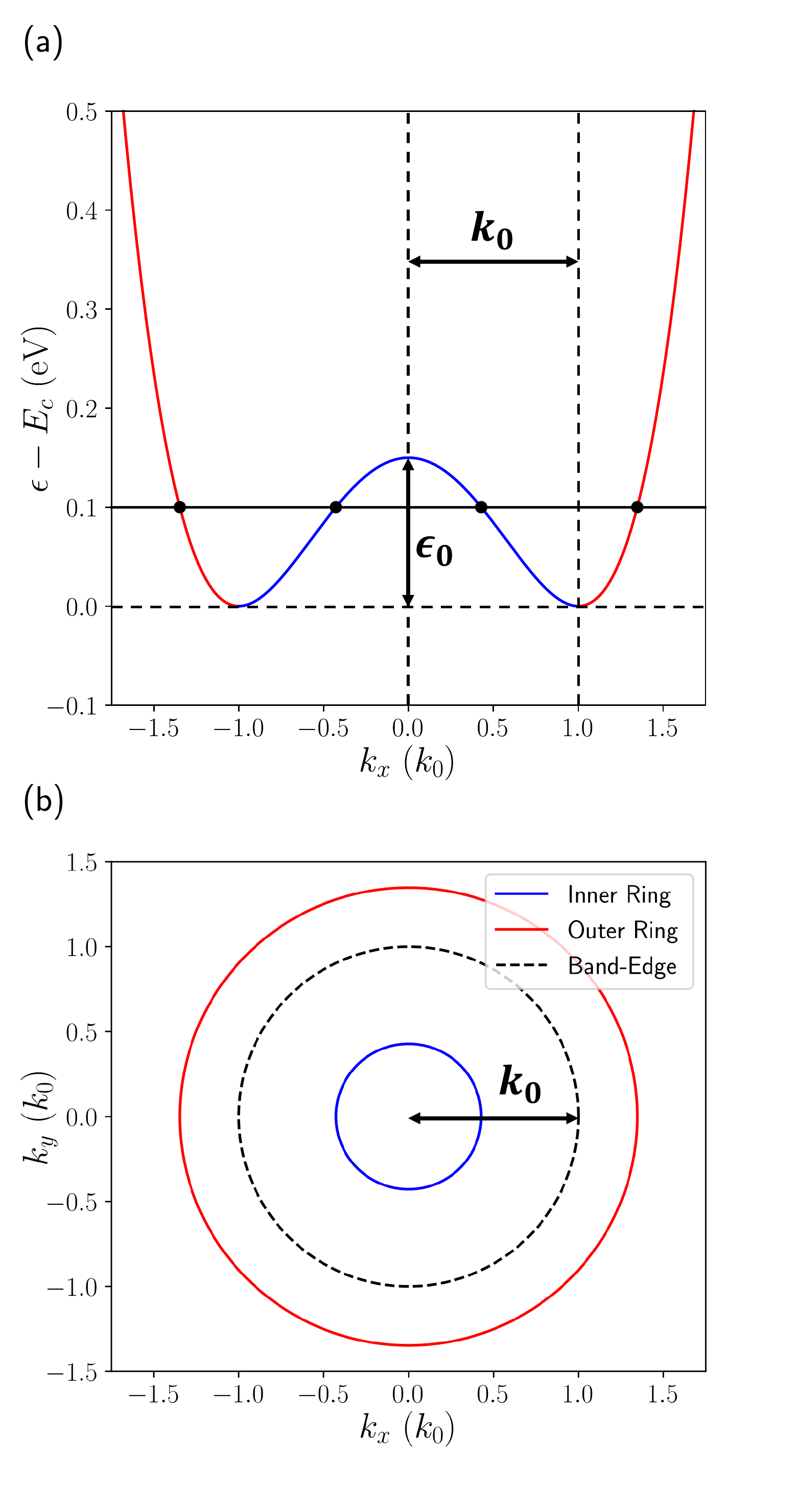}
	\caption{Electron dispersion of a quartic band ($m^*$\,=\,$m_0$ and $\epsilon_0$\,=\,0.15 eV). (a) $E(k)$ versus $k_x$ for $k_y$\,=\,0. (b) Constant energy surface at $E(k)$\,=\,0.1 eV. Note the existence of two distinct surfaces of constant energy (when $E(k)<\epsilon_0$) -- the band edge appears as a dashed line.} \label{fig:quartic_ek}
\end{figure}

\begin{figure}	
	\includegraphics[width=8.5cm]{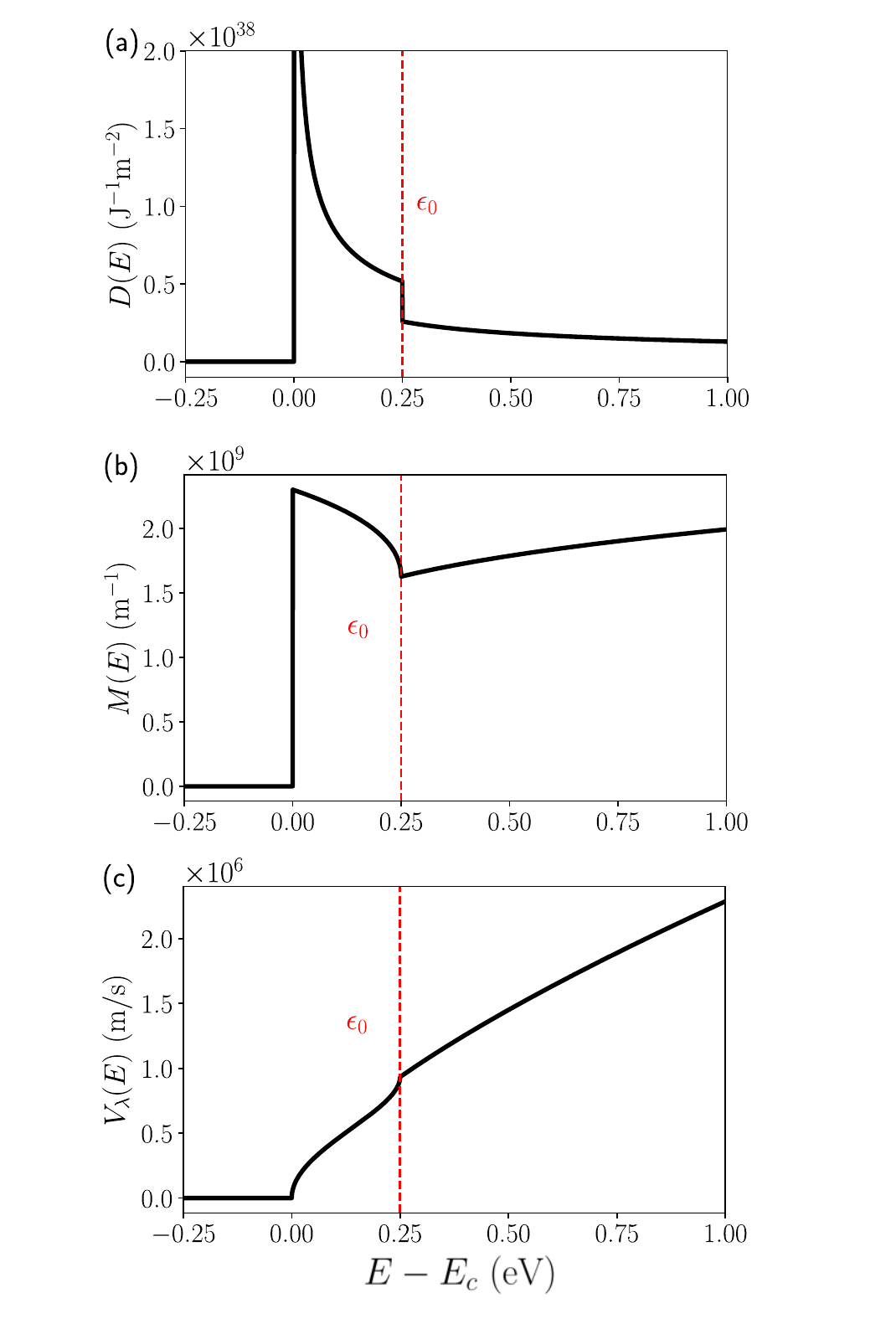}
	\caption{(a) Density-of-states $D(E)$, (b) distribution-of-modes $M(E)$, and (c) average velocity $V_{\lambda}(E)$ versus energy for a quartic band. For these calculations, $m^*$\,=\,$m_0$ and $\epsilon_0$\,=\,0.25 eV (indicated with vertical red line).} \label{fig:quartic_quantities}
\end{figure}

\begin{figure*}	
	\includegraphics[width=17cm]{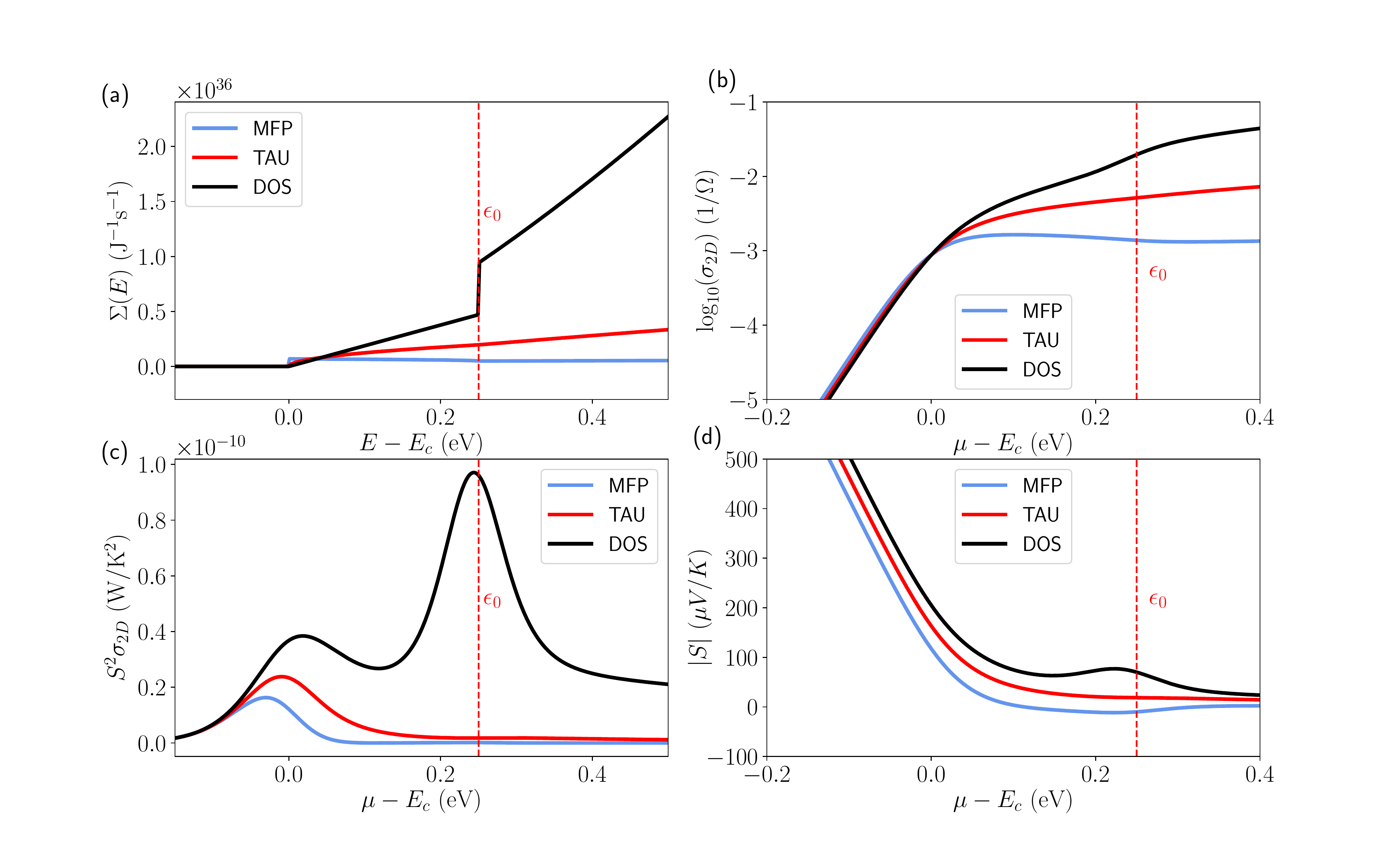}
	\caption{Thermoelectric properties of a quartic band. (a) Transport distribution $\Sigma(E)$ versus energy with $\mu$\,=\,$E_C$. (b) Electrical conductivity $\sigma_{2D}$, (c) power factor $PF$ and (d) magnitude of Seebeck coefficient $|S|$ versus Fermi level $\mu$ using MFP, TAU and DOS scattering models. For these calculations, $m^*$\,=\,$m_0$, $\epsilon_0$\,=\,0.25 eV, $T$\,=\,300~K, and $\langle\langle\lambda\rangle\rangle_{\mu=E_C}$\,=\,10~nm.} \label{fig:quartic_pf}
\end{figure*}

\begin{figure}	
	\includegraphics[width=8cm]{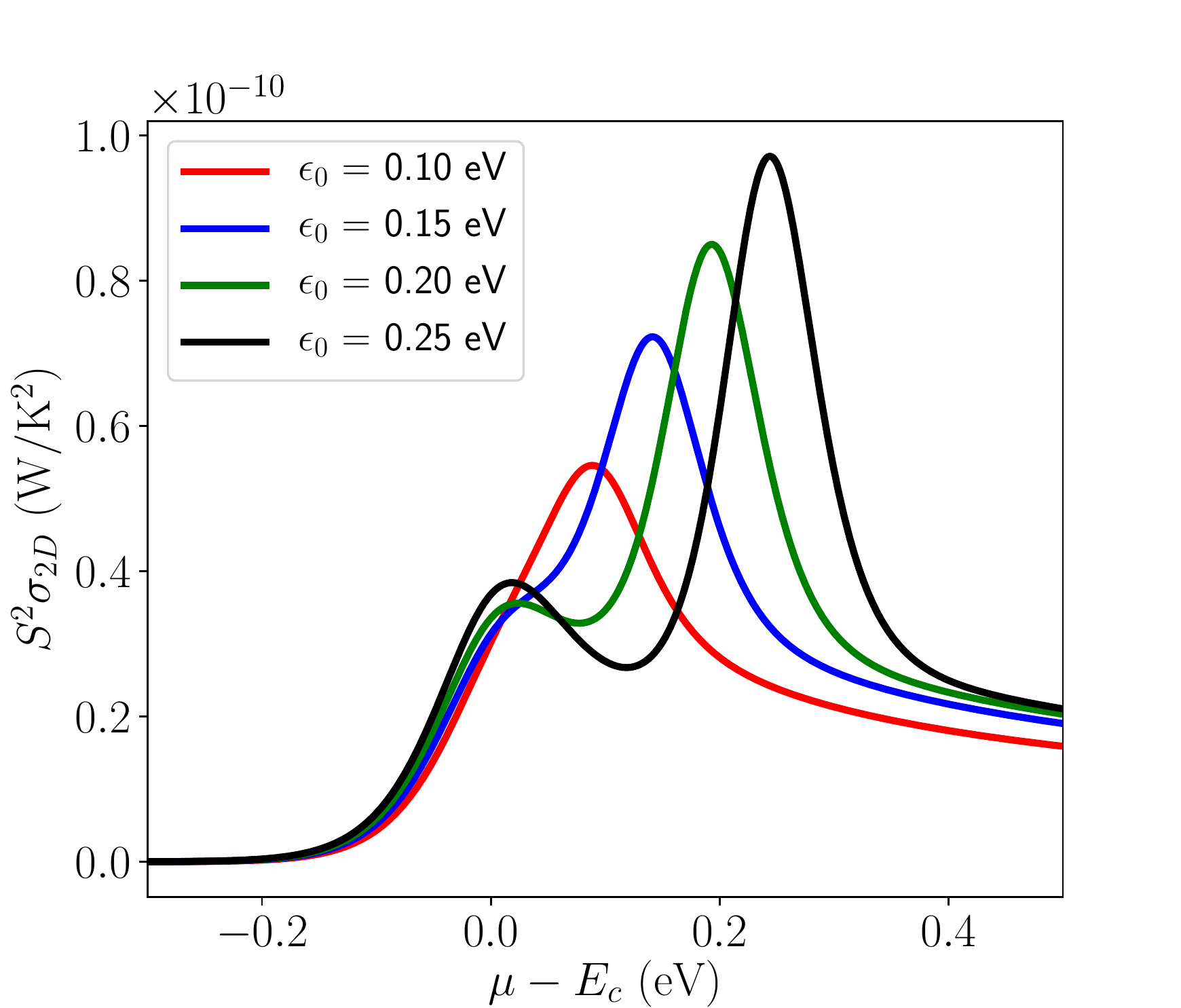}
	\caption{Quartic band. Power factor $PF$ versus Fermi level $\mu$, within the DOS scattering model, for varying $\epsilon$\,=\,0.1, 0.15, 0.20 and 0.25 eV. For these calculations, $m^*$\,=\,$m_0$, $T$\,=\,300~K, and $\langle\langle\lambda\rangle\rangle_{\mu=E_C}$\,=\,10~nm.} \label{fig:quartic_epsilon}
\end{figure}

There are a couple of proposed dispersion models that resemble the more rigorous first-principles computed ring-shaped dispersions, including Rashba band and quartic band models \cite{Wick2015} -- expressions for both are provided in Appendices~\ref{app:quartic}-\ref{app:rashba}. Our analysis will focus on the quartic dispersion, which more closely resembles the DFT-computed dispersions of 2D Bi$_2$Te$_3$ \cite{Zahid2010,Maassen2013a}, one of the early reported ring-shaped band materials for TEs. The dispersion of the quartic model contains a $k^4$ contribution and is given by 
\begin{align}
	\epsilon(k) = \epsilon_0 - \frac{\hbar^2 k^2}{2m^*} + \frac{1}{4\epsilon_0}\left(\frac{\hbar^2 k^2}{2m^*} \right)^2, \label{eq:quartic} 
\end{align}
where $\epsilon_0$ is the energy at $k$\,=\,0 (referred to as the $\Gamma$ point), and $m^*$ is the corresponding effective mass (see Fig.~\ref{fig:quartic_ek}). Note that Eq.~(\ref{eq:quartic}) is for a 2D band structure, with $k$ residing in the $k_x$-$k_y$ plane. This functional form can be derived explicitly by applying a tight-binding model to group-VA elements forming 2D hexagonal lattices, which has been confirmed by DFT \cite{Sevincli2017}. This electronic dispersion has two key features. Firstly, as mentioned above, the band edge consists of a ring of critical points of radius $k_0 = 2\sqrt{m^* \epsilon_0}/\hbar$, rather than a single critical point at $\Gamma$, as in the cases of parabolic or Kane bands. Secondly, the inner ring of states vanishes for energies above $\epsilon_0$. These features are key to understanding the electronic quantities of interest, as discussed next.

The $D(E)$, $M(E)$, and $V_{\lambda}(E)$ are shown in Fig.~\ref{fig:quartic_quantities}, with their respective expressions provided in Appendix~\ref{app:quartic}. The ring of states at the band edge results in a DOM that turns on like a step function. This discrete increase and large number of modes near the band edge is the reason why materials with ring-shaped dispersions were identified as potential high-performance thermoelectrics -- this feature is shared with both quartic and Rashba bands, but not with parabolic or Kane bands. However, there is a second important electronic feature when considering ring-shaped $E(k)$ materials, and that is the discontinuity in the DOS at $\epsilon_0$. Near the $\Gamma$ point, the inner section of the band is roughly that of an inverted 2D parabolic band. Since the DOS of such a band is constant, the ``turning off'' of this portion of the band results in a discontinuous drop in the DOS, as shown in Fig.~\ref{fig:quartic_quantities}. No such discontinuity exists with the Rashba band model; see Appendix~\ref{app:rashba}.

This discontinuity has important consequences for the transport distribution of quartic band materials, as shown in Fig.~\ref{fig:quartic_pf}. Whereas the MFP and TAU models are blind to any changes in the DOS, the DOS scattering model predicts a significant decrease in scattering rate for electrons with energies greater than $\epsilon_0$. This abrupt decrease in scattering rate has been explicitly calculated for the case of charged impurity scattering in warped band materials \cite{Das2019}. The reduced scattering leads to a step-like increase in the transport distribution, not at the band edge, but at $E$\,=\,$\epsilon_0$. As a result, we observe enhancements in both the conductivity and the Seebeck coefficient when the Fermi level is near this feature, for the same reasons described in the parabolic band section (greater magnitude and asymmetry in $\Sigma(E)$). The combined effect of these enhancements is to produce a second local maximum in the power factor; a feature completely overlooked by the MFP and TAU scattering models. This suggests that materials with ring-shaped dispersions may be even better thermoelectric materials than previously believed.

The observation of a second peak in $PF$, with the DOS model, raises the following question: What is the optimal value of $\epsilon_0$, if our goal is to maximize the power factor? Figure~\ref{fig:quartic_epsilon} shows the power factor versus Fermi level, in the case of DOS scattering, for varying $\epsilon_0$. We can see that for sufficiently small $\epsilon_0$, there is only a single $PF$ peak, which splits into two maxima (one near the band edge and one near $\epsilon_0$) as $\epsilon_0$ increases. The maximum power factor is found to increase with $\epsilon_0$, so in principle a larger $\epsilon_0$ should bring about better thermoelectric performance. However, in reality, this will be limited by how far the Fermi level can be pushed into the band, via electrostatic gating or doping. Excessive doping may also begin to alter the electronic structure, which could limit projected performance.

\section{Discussion}
\label{sec:discussion}
The DOS model informs us that good TE materials should have a large distribution-of-modes and velocity, but low density-of-states -- and preferably an abrupt change in these quantities. As illustrated with the quartic dispersion, finding or designing band structures with a rapid decrease in DOS, while maintaining a large DOM and velocity, is another strategy in the search for better TEs. This is contrary to some studies that have sought out materials with a large, narrow peak in DOS, a strategy presumably routed in seeking out a transport distribution resembling a Dirac delta function \cite{Mahan1996} (the following references further discuss this idea \cite{Zhou2011,Jeong2012}). Putting aside the increase in scattering that would likely arise, a large and sharp DOS is not beneficial unless the states have some reasonable velocity; thus, DOS alone is not a good representative quantity for identifying promising TEs. Within the DOS scattering model, the transport distribution is \cite{Witkoske2017}
\begin{align}
\Sigma_{\rm DOS}(E) &= \frac{1}{K_0}\,\frac{\sum_{k,s}v_x^2(k) \,\delta(E-\epsilon(k))}{\sum_{k,s}\delta(E-\epsilon(k))} \nonumber \\
&= \langle v_x^2(E) \rangle/K_0, \nonumber
\end{align}
where we used the relation $M(E)$\,=\,$(h/4)\,\langle |v_x(E)|\rangle\,D(E)$ \cite{Jeong2010,Lundstrom2013}. This expression tells us that, within the DOS model, the most important quality is having a dispersion with large velocities along the transport direction (and preferably an abrupt change at a certain energy).

As mentioned above, while the physical meaning of the transport distribution is not so straightforward, we find it to be the most meaningful {\it single} quantity for assessing a materials potential as a TE. To achieve a large power factor, $\Sigma(E)$ must i) have a large magnitude and ii) be highly asymetric, to preferentially facilitate electron transport above (or below) the Fermi level. Determining why the transport distribution has a particular shape can be easily done by examining its constituent parts, namely the DOM and mean-free-path for backscattering, which have clear interpretations.

This analysis has shown that the choice in scattering model can lead to qualitatively different results, when using analytical electron dispersions. While much effort can go into obtaining numerically accurate dispersions, our findings suggest that the value of such rigorous calculations may be significantly undercut if the treatment of scattering does not capture the correct physics. Researchers investigating the same material, and making use of the same experimental conductivity data to fix their free parameters, may draw very different conclusions about the TE properties if they make use of different scattering approximations. One potential approach to estimate the confidence of a given scattering model could be to compare the results obtained with all three DOS, MFP and TAU models -- if they are similar one can have a high degree of confidence in the results, however if there are significant differences further study would be needed (although the DOS model calculations would likely be most reliable). Thus, if detailed scattering calculations are not feasible, then developing and using reasonably accurate and simple scattering models is important for obtaining reliable results from theoretical modeling.

\section{Conclusions}
\label{sec:conclusions}
Recent first-principles scattering calculations have revealed that the scattering rates often follow the electron density-of-states. Based on this observation, this study has focused on understanding how a simple DOS scattering model, that assumes the scattering rates are proportional to the DOS, influences the calculated thermoelectric properties compared to the more commonly adopted constant scattering-time and constant mean-free-path models (referred to as TAU and MFP models, respectively). Three example analytical dispersions were chosen to illustrate the behavior of the different scattering models, namely a parabolic band (1D, 2D, 3D), a Kane band (1D, 2D, 3D) and a ``ring-shaped'' quartic band (2D). Our findings show that the calculated TE characteristics can vary significantly depending on the choice of scattering model.

The TAU model is found to be inherently more optimistic about TE performance than the MFP model, which has important consequences when predicting the properties of new materials or analyzing experimental results. No such simple trend exists with the DOS model, which can be optimistic or pessimistic depending on the case. Importantly, the scattering models can disagree about whether one can expect benefits from certain dispersions. For example, with a Kane band, the MFP model predicts improved TE performance as the band becomes more ``linear'' (increasing $\alpha$), however the DOS model gives worse performance compared to a parabolic band since larger $\alpha$ increases the DOS and hence the scattering rates -- a feature that both the MFP and TAU models are blind to. The behavior of the different scattering models is explained by analyzing the shape of the transport distribution, which depends on the following physically-transparent quantities: distribution-of-modes, average velocity along transport direction and density-of-states.

The noteworthy cases where the DOS model yields the highest power factor includes the 1D parabolic band and the 2D quartic band. The 1D parabolic band shows better performance with the DOS model, compared to the MFP and TAU models, since the DOS is a decreasing function in energy ($1/\sqrt{E}$) which reduces scattering for high-energy electrons. With the``ring-shaped'' quartic band significantly higher power factors are achieved due to an abrupt \textit{decrease} in the DOS above the band edge, resulting in a sharp decrease in scattering and sharp rise in the transport distribution. This observation suggests a possible strategy for engineering band structures based on identifying or designing materials that exhibit similar drop-offs in DOS (while maintaining large number of modes and velocities), as a potential avenue for future thermoelectric research.

\begin{acknowledgments}
This work was partially supported by DARPA MATRIX (Award No. HR0011-15-2-0037) and NSERC (Discovery Grant RGPIN-2016-04881). C. R. acknowledges support from an NSERC Canada Graduate Scholarship and Nova Scotia Graduate Scholarship.  
\end{acknowledgments}

\appendix

\section{Anisotropic parabolic, effective mass band (1D, 2D, 3D)}
\label{app:para}
For a 1D parabolic band with electronic dispersion given by
\begin{align}
\epsilon(k_x) = \frac{\hbar^2}{2m_x} k_x^2,
\end{align}
we have the following properties: 
\begin{align}
D(E)  &=  \frac{1}{\hbar \pi} \sqrt{\frac{m_x}{2E}}, \\
M(E) &=  \Theta(E), \\
\langle |v_x(E)| \rangle &= \sqrt{\frac{2E}{m_x}}, \\ 
\langle v_x^2(E) \rangle  &= \frac{2E}{m_x}, \\ 
V_\lambda(E) &= 2\sqrt{\frac{2E}{m_x}},
\end{align}
where the following definitions were used $\langle X(E) \rangle = \sum_{k,s} X(k) \delta(E-\epsilon(k))/\sum_{k,s} \delta(E-\epsilon(k))$, and $V_\lambda(E) = 2 \langle v_x^2(E) \rangle/\langle |v_x(E)| \rangle$.

For a 2D parabolic band with electronic dispersion given by
\begin{align}
\epsilon(k_x, k_y) = \frac{\hbar^2}{2} \left(\frac{k_x^2}{m_x} +\frac{k_y^2}{m_y}\right),
\end{align}
we have the following properties: 
\begin{align}
D(E) &=  \frac{\sqrt{m_x m_y}}{\pi \hbar^2}, \\
M(E) &=  \frac{\sqrt{2 m_y E}}{\pi \hbar}, \\
\langle |v_x(E)| \rangle &= \left(\frac{2}{\pi}\right) \sqrt{\frac{2E}{m_x}}, \\
\langle v_x^2(E) \rangle  &= \left(\frac{1}{2}\right) \frac{2E}{m_x}, \\
V_\lambda(E) &= \left(\frac{\pi}{2}\right) \sqrt{\frac{2E}{m_x}}.
\end{align}

For a 3D parabolic band with electronic dispersion given by 
\begin{align}
\epsilon(k_x, k_y, k_z) = \frac{\hbar^2}{2} \left(\frac{k_x^2}{m_x} +\frac{k_y^2}{m_y} +\frac{k_z^2}{m_z}\right),
\end{align}
we have the following properties:
\begin{align}
D(E) &= \frac{\sqrt{2E}}{\pi^2\hbar^3}\sqrt{m_x m_y m_z}, \\
M(E) &= \frac{\sqrt{m_y m_z}}{2 \pi \hbar^2} E, \\
\langle |v_x(E)| \rangle &= \left(\frac{1}{2}\right) \sqrt{\frac{2E}{m_x}}, \\
\langle v_x^2(E) \rangle &=  \left(\frac{1}{3}\right) \frac{2E}{m_x}, \\
V_\lambda(E) &= \left(\frac{4}{3}\right) \sqrt{\frac{2E}{m_x}}.
\end{align}

\section{Kane bands (1D, 2D, 3D)}
\label{app:kane}
For a 1D Kane band with electronic dispersion given by
\begin{align}
\epsilon(1+\alpha \epsilon) = \frac{\hbar^2}{2m^*} k_x^2,
\end{align}
we have the following properties: 
\begin{align}
D(E) &=  \frac{1}{\hbar \pi} \sqrt{\frac{m^*}{2}} \frac{1 + 2 \alpha E}{\sqrt{E(1+\alpha E)}}, \\
M(E) &=  \Theta(E), \\
\langle |v_x(E)| \rangle &= \sqrt{\frac{2E}{m^*}}\frac{\sqrt{1+\alpha E}}{1+2\alpha E}, \\
\langle v_x^2(E) \rangle &= \frac{2E}{m^*}\frac{(1+\alpha E)}{(1+2\alpha E)^2}, \\
V_\lambda(E) &= 2\sqrt{\frac{2E}{m^*}} \frac{\sqrt{1+\alpha E}}{1+ 2\alpha E}.
\end{align}

For a 2D Kane band with electronic dispersion given by 
\begin{align}
\epsilon(1+\alpha \epsilon) = \frac{\hbar^2}{2m^*} \left(k_x^2 + k_y^2\right),
\end{align}
we have the following properties:
\begin{align}
D(E) &= \frac{m^*}{\pi \hbar^2} (1 + 2\alpha E), \\
M(E) &=  \frac{\sqrt{2 m^*}}{\pi \hbar} \sqrt{E(1+\alpha E)}, \\
\langle |v_x(E)| \rangle &=  \left(\frac{2}{\pi}\right) \sqrt{\frac{2E}{m^*}} \frac{\sqrt{1+\alpha E}}{1+2\alpha E}, \\
\langle v_x^2(E) \rangle &=  \left(\frac{1}{2}\right) \frac{2E}{m^*}\frac{(1+\alpha E)}{(1+2\alpha E)^2}, \\
V_\lambda(E) &= \left(\frac{\pi}{2}\right) \sqrt{\frac{2E}{m^*}} \frac{\sqrt{1+\alpha E}}{1+2\alpha E}.
\end{align}

For a 3D Kane band with electronic dispersion given by 
\begin{align}
\epsilon(1+\alpha \epsilon) = \frac{\hbar^2}{2m^*} \left(k_x^2 + k_y^2 +k_z^2\right), 
\end{align}
we have the following properties:
\begin{align}
D(E) &=  \frac{\sqrt{2}(m^*)^{3/2}}{\pi^2 \hbar^3} \sqrt{E(1+\alpha E)} (1+2\alpha E), \\
M(E) &= \frac{m^*}{2\pi \hbar^2} E(1+\alpha E), \\
\langle |v_x(E)| \rangle &= \left(\frac{1}{2}\right) \sqrt{\frac{2E}{m^*}} \frac{\sqrt{1+\alpha E}}{1+2\alpha E}, \\
\langle v_x^2(E) \rangle &= \left(\frac{1}{3}\right) \frac{2E}{m^*} \frac{(1+\alpha E)}{(1+2\alpha E)^2}, \\
V_\lambda(E) &= \left(\frac{4}{3}\right) \sqrt{\frac{2E}{m^*}} \frac{\sqrt{1+\alpha E}}{1+2\alpha E}.
\end{align}

\section{Quartic band (2D)}
\label{app:quartic}
For a 2D ring-shaped, quartic band with electronic dispersion given by 
\begin{align}
\epsilon(k_x, k_y) = &\epsilon_0 - \frac{\hbar^2}{2m^*}\left(k_x^2 + k_y^2\right) \nonumber\\ 
&+ \frac{1}{4\epsilon_0} (\frac{\hbar^2}{2m^*})^2\left(k_x^2 + k_y^2\right)^2, 
\end{align}
we have the following properties:
\begin{align}
 D(E) = \begin{cases}
\frac{2m^*}{\pi \hbar^2}\sqrt{\frac{\epsilon_0}{E}}, & E < \epsilon_0 \\
\frac{m^*}{\pi \hbar^2}\sqrt{\frac{\epsilon_0}{E}}, & E > \epsilon_0
\end{cases}
\end{align}
\begin{align}
&M(E) = \nonumber\\
&\begin{cases}
\frac{2\sqrt{m^* \epsilon_0}}{\pi \hbar}\left(\sqrt{1+\sqrt{\frac{E}{\epsilon_0}}} +\sqrt{1-\sqrt{\frac{E}{\epsilon_0}}} \right), & E < \epsilon_0 \\
\frac{2\sqrt{m^* \epsilon_0}}{\pi \hbar}\sqrt{1+\sqrt{\frac{E}{\epsilon_0}}} , & E > \epsilon_0
\end{cases}
\end{align}
\begin{align}
&\langle |v_x(E)| \rangle = \nonumber\\ 
&\begin{cases}
\frac{2}{\pi} \sqrt{\frac{E}{m^*}}\left(\sqrt{1+\sqrt{\frac{E}{\epsilon_0}}} +\sqrt{1-\sqrt{\frac{E}{\epsilon_0}}} \right), & E < \epsilon_0 \\
\frac{4}{\pi} \sqrt{\frac{E}{m^*}}\sqrt{1+\sqrt{\frac{E}{\epsilon_0}}} , & E > \epsilon_0
\end{cases}
\end{align}
\begin{align}
&\langle v_x^2(E) \rangle = \nonumber\\ 
&\begin{cases}
2\frac{E}{m^*} , & E < \epsilon_0 \\
2\frac{E}{m^*}\left(1+\sqrt{\frac{E}{\epsilon_0}} \right) , & E > \epsilon_0
\end{cases}
\end{align}
\begin{align}
& V_\lambda(E) = \nonumber \\
&\begin{cases}
2\pi\sqrt{\frac{E}{m^*}}\frac{1}{\sqrt{1+\sqrt{\frac{E}{\epsilon_0}}} +\sqrt{1-\sqrt{\frac{E}{\epsilon_0}}}} , & E < \epsilon_0 \\
\pi \sqrt{\frac{E}{m^*}}\sqrt{1+\sqrt{\frac{E}{\epsilon_0}}}. & E > \epsilon_0
\end{cases}
\end{align}

\section{Rashba band (2D)}
\label{app:rashba}
\begin{figure}
	\includegraphics[width=9cm]{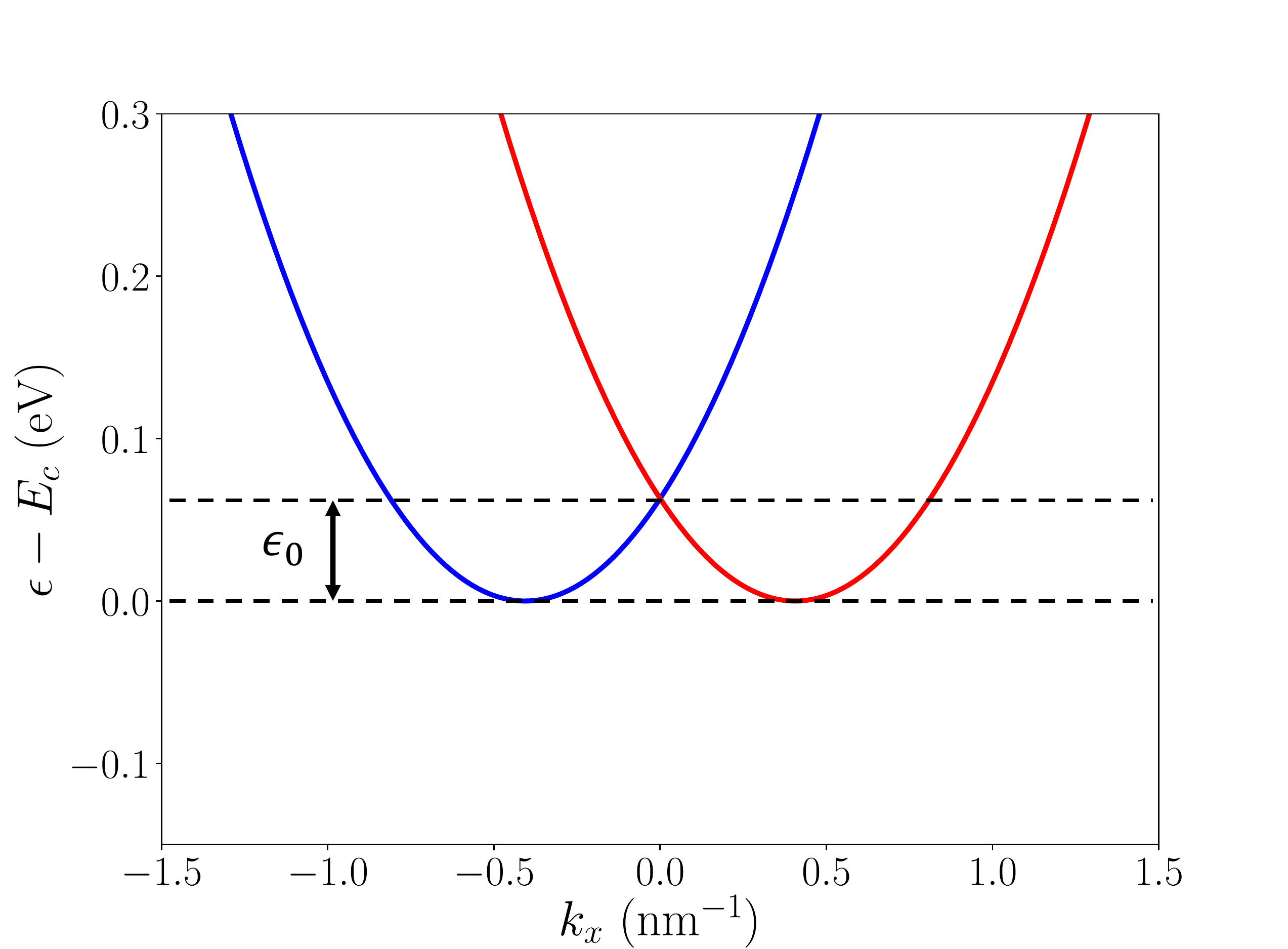}
	\caption{Electron dispersion of a Rashba band ($m^*$\,=\,0.1\,$m_0$ and $\alpha_R$\,=\,5$\times$$ 10^{-29}$ J-m).}   \label{fig:2D_rashba}
\end{figure}
For a 2D ring-shaped, Rashba band with electronic dispersion given by 
\begin{align}
\epsilon(k_x, k_y) = \epsilon_0 + \frac{\hbar^2}{2m^*}\left(k_x^2 + k_y^2\right) \pm \alpha_R \sqrt{k_x^2 + k_y^2}, 
\end{align}
where 
\begin{align}
	\epsilon_0 = \frac{\alpha_R^2 m^*}{2 \hbar^2}, 
\end{align}
we have the following properties:
\begin{align}
D(E) &= \begin{cases}
\frac{m^*}{\pi \hbar^2}\sqrt{\frac{\epsilon_0}{E}}, & E < \epsilon_0 \\
\frac{m^*}{\pi \hbar^2}, & E > \epsilon_0
\end{cases} \\
M(E) &= \begin{cases}
\frac{\sqrt{2m^*\epsilon_0}}{\pi \hbar}, & E < \epsilon_0 \\
\frac{\sqrt{2m^*E}}{\pi \hbar}, & E > \epsilon_0
\end{cases} \\
\langle |v_x(E)| \rangle &= \left(\frac{2}{\pi}\right)\sqrt{\frac{2E}{m^*}}, \\
\langle v_x^2(E) \rangle &= \left(\frac{1}{2}\right)\frac{2E}{m^*}, \\
V_\lambda(E) &= \left(\frac{\pi}{2}\right)\sqrt{\frac{2E}{m^*}}. 
\end{align}


\end{document}